\documentclass[runningheads,12pt]{llncs}
\usepackage{url} % takes care of hyperlinks, preferred over hyperref
\usepackage[ruled,lined]{algorithm2e}% provides Algorithm environment
\usepackage{graphicx}% allows for inclusion of graphic files (figures)

\usepackage{amssymb}
\usepackage{amsmath}
\usepackage{array}
\usepackage{enumitem}
\usepackage{eurosym}
\usepackage{mymacros}
\usepackage{wasysym}
\usepackage{pseudocode}
\usepackage{color}

\def\mi#1{\mathit{#1}}
\def\univ{{\cal U}}
\def\Uev{\univ_{\mi{ev}}}
\def\Uact{\univ_{\mi{act}}}
\def\Utime{\univ_{\mi{time}}}
\def\Ucase{\univ_{\mi{case}}}
\def\Ures{\univ_{\mi{res}}}
\def\Uatt{\univ_{\mi{att}}}
\def\Uval{\univ_{\mi{val}}}
\def\Umap{\univ_{\mi{map}}}
\def\powerset{\mathcal{P}}
\def\bag{\mathcal{B}}
\def\la{\langle}
\def\ra{\rangle}

\def\card#1{\ensuremath{\left\vert{#1}\right\vert}}

\newcommand{\nat}{\mbox{I\hspace*{-.2em}N}}

\def\pre#1{\ensuremath{\bullet{#1}}}
\def\post#1{\ensuremath{{#1}\kern-.05ex\bullet}}
\def\en#1{\ensuremath{\protect{[{#1}\rangle}}}

\begin{document}

\title{How to Write Beautiful Process-and-Data-Science Papers?}
\titlerunning{How to Write Beautiful Process-and-Data-Science Papers?}

\author{Wil M.P. van der Aalst}
\authorrunning{Wil van der Aalst}
% First names are abbreviated in the running head.
% If there are more than two authors, 'et al.' is used.
%
\institute{Process and Data Science (PADS), RWTH Aachen University, Germany\\
\url{wvdaalst@pads.rwth-aachen.de} ~~
\url{www.vdaalst.com}
 }
\maketitle              % typeset the header of the contribution

\begin{abstract}
After 25 years of PhD supervision, the author noted typical recurring problems that make papers look sloppy, 
difficult to read, and incoherent. The goal is not to write a paper for the sake of writing a paper, 
but to convey a valuable message that is clear and precise. 
The goal is to write papers that have an impact and are still understandable a couple of decades later.
Our mission should be to create papers of high quality that people want to read and that can stand the test of time.
We use Dijkstra's adagium ``Beauty Is Our Business'' to stress the importance of simplicity, correctness, and cleanness.  
\end{abstract}
\section{Introduction}
\label{sec:introduction}

This paper is not a scientific paper, but a personal ``style guide'' to \textbf{help authors to write better scientific articles}.
Although the scope is not limited to a specific topic and applies to any computer science or information systems paper,
examples are taken from the field of \textbf{Process and Data Science} (PADS).

This ``style guide'' is a \textbf{living document} that will be extended over time and feedback is welcome. 
Note that ``style'' is, of course, a very personal matter. 
However, when working in a group or community, it helps to stick to standard notations and conventions, and ensure a minimum quality level. 
For example, when defining a Petri net, we normally use $N=(P,T,F)$
and there is no need to explain that $P$ are the places, $T$ are the transitions, and $F$ are the flows (i.e., arcs connecting places and transitions). 
Of course, one can define a Petri net as $P=(S,T,I,O)$ or 
$\nu = (P,T,\alpha^+,\alpha^-)$, but now explanations are needed for things that are standard. 
When denoting bags, we can write $[a^2,b^3,c]$ or use a range of other notations, e.g., $\{a \rightarrow 2,a \rightarrow 3, a \rightarrow 1\}$. Some authors use $\subset$ to denote a subset, but most use $\subseteq$ (reserving $\subset$ for strict subsets). 
These examples show that it is helpful to use conventions. 
In science, we are most productive when we are able to \textbf{stand on the shoulders of others}
and use discoveries, insights, and conventions.
Moreover, we want others to stand again on our shoulders, so we should make this as easy as possible.
This style guide aims to support this.

This style guide is not intended to limit \textbf{academic freedom} in any way. 
You can be stubborn if you produce high-quality results.
However, it is not acceptable to use non-standard definitions and terms that have obvious problems,
and then expect others (e.g., readers, reviewers, and supervisors) 
to spend hours trying to understand something simple or even repair things.
This is a waste of energy and leads to a lower overall quality (because it deviates from the essence). 
Using pattern matching, one can quickly spot errors if the same conventions are used by a community. 

Of course, \textbf{style and notations are a matter of taste} and sometimes authors 
refer to other papers in mathematics, machine learning, simulation, engineering, security, etc.\ that use a different style.
Although this may be a valid argument, one should realize that \textbf{many papers have formal or logical problems} 
and were written by inexperienced authors (e.g., starting PhDs). 
Hence, the fact that something is published does not make it correct. 
The number of accepted papers with constructs such as ``$X \setminus \emptyset$'' and 
mixing up things at the type and instance level is shocking.
In many disciplines, it is not necessary to be very precise and the focus is on engineering, societal or economic aspects. 
However, when dealing with data and processes and aiming to conduct top-quality research, one needs to be very precise.

When writing a paper, be prepared to throw it away and start from scratch when notations do not work. 
As Edsger Dijkstra once said: 
``When we recognize the battle against chaos, mess, and unmastered complexity as one of computing science's major callings, we must admit that \textbf{Beauty Is Our Business}''.\footnote{Edsger W. Dijkstra. \emph{Some Beautiful Arguments Using Mathematical Induction}, EWD 697,
 1978.}

The remainder tries to highlight some of the common problems that the author frequently noted over the last 25 years.
The goal is to improve the quality and readability in the field of process and data science by sharing these insights.

\section{How to Start a Paper?}
\label{sec:start}

Very few readers read a paper from beginning to end.
As they say, \textbf{``You only get one chance to make a first impression''} meaning that if it is not clear in 5-10 minutes what the paper is about, the reader will discard it and never look at it again. 
Therefore, the title, abstract, and introduction need to clarify what the paper is about and what is new.
The title should be \textbf{catchy} and, at the same time, \textbf{specific} enough. 
If you use half of the abstract and introduction explaining what process mining is, 
then your real message is likely to get lost.

Use the following checklist when you write the introduction:
\begin{itemize}
\item \textbf{What is the problem and why is the problem relevant?} You need to create a \textbf{sense of urgency}. It is not enough to describe a research area.
\begin{itemize}
 \item If it is a new problem: Why is it relevant? 
 \item  If it is not a new problem: Where do existing approaches fail?
 \end{itemize}
\item \textbf{What is your unique way of approaching the problem?} You need to be able to describe your solution approach (without going into details). You also need to explain why this is novel and interesting. Avoid describing an ``area''.
\item \textbf{What is the input and what is the output?} Most of the things we do have a clear input and output, e.g., for discovery, an event log goes in, and a process model comes out. Do not assume that the reader knows this already. Thinking in terms of input and output makes the paper more concrete. Also, state how you are going to evaluate the output.
\item \textbf{Use at least one diagram in the introduction!} 
An introduction consisting of only ``flat text'' is likely to be skipped. 
There are two types of diagrams you can use: An overview diagram or an example showing schematically input and output. Do not underestimate how a good diagram can create a mental image that sticks.
\end{itemize}

Never use the argument that things will be clear when the reader reads the whole paper. \textbf{This will not happen!}
The core message should be clear in 5-10 minutes.
\textbf{Think of a paper like a six course dinner}: introduction (Antipasti), preliminaries (Primi Piatti), approach (Secondi Piatti), implementation (Contorni), experiments (Dolce), conclusion (Caffe).
You cannot say ``If you do not like the Antipasti and Primi Piatti, just continue eating until the end and you will also appreciate the first courses.'' Your guests will escape as soon as possible and never come back.\footnote{When it comes to food, people may be hungry and stay until the last course. However, when it comes to papers, readers typically leave after one or two courses. Therefore, one needs to be to the point and convince the reader to continue.}

\textbf{Make consistent assumptions about your audience.} Many papers explain elaborately what process mining is and formalize Petri nets, event logs, etc.\ using high-quality definitions. However, the moment the reader reaches the original contribution, 
suddenly the reader is supposed to have ``superpowers'' able to distill complex ideas and formalizations from running text.
Be sure that the paper is \textbf{balanced}. 
Note that there is a natural tendency to elaborately write about the things that are clear and standard (because it is easy) and be brief and vague about the new stuff. Try to resist this.

\textbf{Repetitions do not help.} There is also a tendency to repeat informal statements when formalizations or concepts are not clear enough.
This will make things worse. Remove things you cannot explain adequately.

\section{American Spelling}
\label{sec:spell}

There are spelling differences between American English, British English, Canadian English, and Australian English. These are all correct. However, the author recommends  using \textbf{American English}. This is what the majority of the researchers in the field is using. 
Moreover, the way that non-native speakers write is closer to American English and definitely not similar to British English. 

Hence, let us use ``modeling - analyze - defense - labor - color - organize - program'' rather than ``modelling - analyse - defence - labour - colour - organise - programme''. Note that this is easy: Just use a spell checker.

However, there are many errors that \textbf{a spell checker cannot capture}. Given your language background, you may drop ``a'' and ``the'' or insert these at places where they do not belong.\footnote{See \url{https://en.wikipedia.org/wiki/Dunglish} for funny Dutch mistakes like ``are a nation of undertakers'', ``make that the cat wise'', ``death or the gladioli'', ``I always get my sin'', ``we have to look further than our nose is long'', ``may I thank your cock for the lovely dinner'', `` how do you do your wife'', etc.} You need to be aware of the typical mistakes you make and read the text to remove these problems. You cannot expect others to repair the whole document for you. \textbf{If someone points out such a problem in your text, do not just repair the individual sentence!} Look at the whole paper for recurrences and write down such errors, so that you check it again in your next paper. 

These things also apply to slides. Showing slides in meetings with a typo on each slide shows a lack of respect. Of course, it is OK to make mistakes; we all do. However, you need to be self-critical to be taken seriously.

There are many books on typical errors in English. For example, see the list with ``50 Common Grammar Mistakes in English'' created by Rebecca Ezekiel (\url{www.engvid.com}). 

There are many textbooks and websites that point out \textbf{recurring problems}, such as using ``which'' incorrectly. One should always use ``who'' for people, 
``that'' should be used to introduce a restrictive clause (leaving it out changes the meaning) and 
``which'' should be only used to introduce a non-restrictive or parenthetical clause (leaving it out does not change the meaning). Also ``which'' has a comma before and ``that'' not.
Also read up on when to use 
``each'', ``any'',
``few'', ``little'',``fewer'', ``less'', ``many'', ``more'', etc. 

We only use \textbf{capitals} for names and to introduce acronyms.
For example, ``\ldots colored Petri nets and Colored Petri Nets (CPN) \ldots'' rather than 
``\ldots colored petri nets and Colored Petri Nets \ldots''.
Therefore, do not start writing ``Process Mining'' in the middle of a sentence. Most people that do this tend to use capitals randomly.
You can say ``in this figure'', ``in this section'', ``in Figure 5'', ``in Section 2'', ``in Celonis'', and ``in our tool''. However, ``in this Figure'', ``in this Section'', ``in figure 5'', ``in section 2'', ``in the Figure 5'', ``in the Section 2'', ``in the figure 5'', ``in the section 2'', `in the celonis'', and ``in tool'', etc.\ are all \textbf{wrong}. 

There are many more subtle things like when to use a comma, to avoid using ``on the other hand'' without the ``on the one hand'', and know that words like semantics are plural. 
It is impossible to point out all common mistakes here.

In a scientific paper, one never uses ``I'' and ``you''. One can use ``we'', but the third person is generally used in scientific writing (``our approach was implemented in ProM'', ``experimentation showed'', ``it was assumed that'', etc. The author prefers to have a balance between ``we'' and the third person.

Finally, it is highly recommended to \textbf{use short sentences}. Use sentences of \textbf{less than 30 words}. In Springer LNCS format, a sentence should never exceed two lines.  Native speakers can formulate beautiful longer sentences. However, if you are not a native speaker and/or your audience is composed of many non-native speakers, keep it simple. Why make things difficult for yourself and/or your audience?

\section{Multisets, Sequences, Etc.}
\label{sec:multi}

As mentioned, authors working in a particular field should try and use a \textbf{uniform style and uniform notations}. 
For the people working in process and data science, it makes sense to use the Process Mining book \cite{process-mining-book-2016}, the DADP paper \cite{aal_decomp_procmin_generic_dapd}, or the first two chapters of the Process Mining Handbook \cite{PMhandbook-SS22} as a reference.
Of course, there may be good reasons to use different notations, but let's avoid using a plethora of notations without a clear reason.

Some examples from \cite{aal_decomp_procmin_generic_dapd}:
 \begin{itemize}
 \item $\bag(A)$ is the set of all multisets over some set $A$. An example bag is $[x^3,y^2,z]$. $[~]$ is the empty multiset. Note that technically $\bag(A) = A \rightarrow \nat$, i.e., $[x^3,y^2,z] = \{ (x,3),(y,2),(z,1)\}$, but we avoid using this notation as much as possible.
 \item $\sigma = \langle a_1,a_2, \ldots, a_n\rangle \in X^*$ denotes a sequence over $X$ of length $n$. $\langle~\rangle$ is the empty sequence.
 \item $\sum_{x \in \langle a,a,b,a,b\rangle} f(x) = \sum_{x \in [a^3,b^2]} f(x) = 3 f(a) + 2 f(b)$.
 \item $[f(x) \mid x \in \langle a,a,b,a,b\rangle] = [f(x) \mid x \in [a^3,b^2]] = [(f(a))^3,(f(b))^2]$.
 \item $\{ x \in [a^3,b^2]\} = \{ x \in  \langle a,a,b,a,b\rangle\} = \{a,b\}$.
 \item $f \in X \not\rightarrow Y$ is a partial function with domain $\mi{dom}(f) \subseteq X$ and 
 range $\mi{rng}(f) = \{f(x)\mid x\in \mi{dom}(f)\} \subseteq Y$.
$f \in X \rightarrow Y$ is a total function, i.e., $\mi{dom}(f) = X$.
 \item $f = \{(a,2),(b,3),(c,4)\}$ can be viewed as a function with $\mi{dom}(f) \allowbreak = \{a,b,c\}$ and $\mi{rng}(f) = \{2,3,4\}$.
 $f' = f\oplus\{(c,6),(d,7)\} = \{(a,2),\allowbreak (b,3),\allowbreak (c,6),\allowbreak (d,7)\}$ updates $f$, e.g., $f'(a)=2$, $f'(c)=6$, $f'(d)=7$.
\item $f\tproj_{Q}$ is the function projected on $Q$: $\mi{dom}(f\tproj_{Q}) = \mi{dom}(f) \cap Q$ and $f\tproj_{Q}(x) = f(x)$ for $x \in \mi{dom}(f\tproj_{Q})$.
\item $N=(P,T,F)$  is a Petri net, $M \in \bag(P)$ is a marking, $\pre x$ and $\post x$ are pre- and post-sets, etc.
\item Alignments are denoted as $\gamma = \begin{array}{|c|c|c|c|c|}
a & b  & c & \nomove & e \\ \hline
a & \nomove & c & d & e  \\
\end{array}$ or 
$ \gamma = \begin{array}{|c|c|c|c|c|c|}
a & b & \nomove & c & \nomove & e \\ \hline
a & \nomove & \tau & c & d & e  \\
t1 & & t3 & t4 & t5 & t6   \\
\end{array}$. The top row should refer to the event log and the bottom row to the process model.
We need to distinguish transition names from activities in a labeled Petri net.
 \end{itemize}
 
\noindent Some conventions:
\begin{itemize}
 \item Start counting with $1$. For ``religious reasons'' some people prefer to write $0 \leq i \leq n-1$ rather than $1 \leq i \leq n$. 
However, calling the first element the ``zero-th element'' is confusing for most.\footnote{Although some prefer to start with 0, see for example Edsger W. Dijkstra's  \emph{Why Numbering Should Start at Zero}, EWD-831, 1982.}
 \item Use small letters for individual elements, e.g., $a$, $b$, $p$, $t$, $x$, $y$, etc.
 \item Use capital letters for sets, e.g., $A$, $P$, $T$, etc.
 \item Use lowercase letters or short lowercase words for functions, e.g., $f$, $g$, $\mi{min}$, $\mi{mean}$, etc.
 \item For multisets, this is not so clear since they can be viewed as a special kind of set or as a special kind of function. 
 \item Try to combine uppercase and lowercase to reduce the cognitive load, e.g., $a \in A$, $p \in P$, and $t \in T$ is easier to read and remember than 
 $x \in A$, $p \in Y$, and $z \in T$.
 \item Use $\sigma$ (and if needed $\rho$ and $\gamma$) for sequences.
 \item Use $i$, $j$, $k$, $n$, $m$ for integers, e.g., $1 \leq i < j \leq n$.
 \item Avoid excessive use of Greek symbols and use short names for functions, e.g., $\mi{min}(A)$, $\mi{first}(\sigma)$, $\mi{sort}(Q)$ are easier to remember than  $\theta(A)$, $\vartheta(\sigma)$, and $\Theta(Q)$. Of course there are Greek symbols like $\sigma$ (for sequence), $\tau$ (for silent activities), and $\Delta$ (for difference) that have a standard meaning and can be used. 
 \item Use $\exists_{x \in X}\ b(x)$,  $\forall_{x \in X}\ b(x)$, $\neg b$, $b_1 \ \wedge \ b_2$, $b_1 \ \vee \ b_2$,
 $b_1 \ \Rightarrow \ b_2$, and do not mix these notations with ``and'', ``if'', or a comma as a conjunction.
\end{itemize}

Explicitly consider the \textbf{binding} of each variable, e.g., $\{x \in X \mid \exists_{y \in Y} f(x,y) = z\}$ is incorrect unless $z$ has a constant value in this context. For each variable, check where it is bound. $f(x,y) = \frac{x+y}{z}$ is incorrect unless $z$ is a constant, i.e., the right-hand side needs to fully depend on the left-hand side. Hence, $\mi{freq}(a)$ cannot magically depend on an event log $L$. Write something like $\mi{freq}_L(a)$ if it does.

If formalizations are not your natural habitat, then print your paper and draw arcs between concepts and variables used in formalizations to the locations where they are introduced. Moreover, \textbf{create ``instances'' of the mathematical objects you define}.
For example, when using $N=(P,T,F,l)$ with $F \subseteq (P \times T) \cup (T \times P)$ and $l \in T \not\rightarrow \Uact$, \textbf{force yourself} to write instances like $P = \{p1,p2, \ldots,p8\}$, $T = \{t1,t2, \ldots t6\}$, 
$F = \{(p1,t1),\allowbreak$ $(t1,p2),\allowbreak$ $(t1,p3),\allowbreak$  $(t1,p4),\allowbreak$
$(p2,t2),\allowbreak$ $(p3,t3),\allowbreak$ $(p4,t4),\allowbreak$
$(t2,p5),\allowbreak (t3,p6),\allowbreak$ $(t4,p7),\allowbreak$ 
$(p7,t5),\allowbreak$ $(t5,p4),\allowbreak$
$(p5,t6),\allowbreak (p6,t6),\allowbreak (p7,t6),\allowbreak 
(t6,p8\}$ and $l = \{(t1,a),\allowbreak (t2,b),\allowbreak (t3,c),\allowbreak (t4,d)\}$.

Some additional examples to illustrate the need to create concrete instances of formal expressions. 
Let $x \in \powerset(\powerset({\Uact}^*))$, $y \subseteq \bag({\Uact}^*)$, and $z = \powerset(\powerset(\emptyset))$. 
What are example values for $x$, $y$, and $z$? Here are some
$x = \{\{\la \ra,\la a,b \ra\},\allowbreak \{\la \ra,\la b,a \ra\}\}$, $x = \emptyset$, $x = \{\emptyset\}$,
$y = \{ [\la \ra^2,\la a,b \ra^3],\allowbreak [\la \ra^3,\la a,b \ra^2]\}$, $y = \emptyset$, $y = \{[\,]\}$, and
$z= \{\{\emptyset\},\emptyset\}$.
\textbf{Yes, this is tedious, but it helps you to detect formalization errors and think of corner cases.}
 
\section{Definitions and Universes}
\label{sec:defs}

Whenever you write something, \textbf{the scope should be clear}. You cannot define some artifact (e.g., a footprint matrix) in running text and then assume you can ``access'' it whenever you want.

There are two ways to introduce something: (1) \textbf{using a ``universe'' construct}  and (2) \textbf{using the ``Let'' statement}.
We can say that $\nat$ is the set of natural numbers and use it throughout the paper without a ``Let'' statement, i.e., 
it holds universally and does not need to be declared.
We can also say that $\univ_{\mi{act}}$ is the universe of activities and $E$ is the universe of activities. However, this means that these cannot change. They cannot be used to refer to a specific set of activities and events.
If you plan to modify attributes of events or plan to change the network structure, you cannot use the ``universe'' construct.

\textbf{Each definition needs to be self-contained.} The fact that you introduced artifacts before or computed intermediate results does not mean that you can assume their presence implicitly. You need to ``wire'' the definitions explicitly using ``Let'' statements. Consider, for example, 
the following toy definition.

\begin{definition}[Example]\label{def:ex}
Let $L \in \bag(A^*)$ be an event log over a set of activities $A$
and let $N=(P,T,F,l)$ be a labeled Petri net.
$L$ and $N$ are compatible if and only if $\cup_{\sigma \in L} \{a \in \sigma\} = \{l(t) \mid t \in \mi{dom}(l)\}$.
\end{definition}

You cannot drop the first line with the two ``Let'' statements, even when you talked about logs and Petri nets before.
We need to know that $L$ is an event log (i.e., a multiset of sequences over a set of activities)
and that $N$ is a labeled Petri net with transitions $T$ and a labeling function $l$.
The \textbf{context has to be perfectly clear}. 
The paper may introduce multiple types of event logs and Petri nets, e.g., filtered event logs and short-circuited workflow nets. This may sound trivial, but note that $\pre t$ assumes the presence of a specific Petri net.
In many papers, it is not always clear to which Petri net $\pre t$ refers.

As an example, we show three ways to introduce an event log where each event refers to a case, activity, timestamp, and resource. Assume we already introduced $\Ucase$ as the universe of cases, $\Uact$ as the universe of activities, 
$\Utime$ as the universe of timestamps, and $\Ures$ as the universe of resources.

\begin{definition}[Approach 1]\label{def:evenlog1}
An event $e$ is tuple $e=(c,a,t,r) \in \Ucase \times \Uact \times \Utime \times \Ures$ 
referring to case $c$, activity $a$, timestamp $t$, and resource $r$ of event $e$.
An event log $L$ is a multiset of events, i.e., $L \in \bag(\Ucase \times \Uact \times \Utime \times \Ures)$.
\end{definition}

\begin{definition}[Approach 2]\label{def:evenlog2}
$\Uev$ is the universe of events. $e \in \Uev$ is an event,
$\pi_{case}(e) \allowbreak \in \allowbreak \Ucase$ is the case of $e$,
$\pi_{act}(e) \in \Uact$ is the activity of $e$,
$\pi_{time}(e) \in \Utime$ is the timestamp of $e$, and
$\pi_{res}(e) \in \Ures$ is the resource of $e$. An event log $L$ is a set of events $L \subseteq \Uev$. 
\end{definition}

\begin{definition}[Approach 3]\label{def:evenlog3}
$\Uev$ is the universe of events, 
$\Uatt$ is the universe of attribute names ($\{\mi{case},\mi{act},\mi{time},\mi{res}\} \subseteq \Uatt$),
$\Uval$ is the universe of attribute values, 
and $\Umap = \Uatt \not\rightarrow \Uval$ is the universe of attribute value mappings.
An event log is a tuple $L = (E,\pi)$ with
$E \subseteq \Uev$ as the set of events and $\pi \in E \rightarrow \Umap$
such that for any $e \in E$:
$\{\mi{case},\mi{act},\mi{time},\mi{res}\} \subseteq \mi{dom}(\pi(e))$
and  
$\pi(e)(\mi{case}) \in \Ucase$ is the case of $e$,
$\pi(e)(\mi{act}) \in \Uact$ is the activity of $e$,
$\pi(e)(\mi{time})\in \Utime$ is the timestamp of $e$, and
$\pi(e)(\mi{res}) \in \Ures$ is the resource of $e$. 
\end{definition}

Let us compare the three approaches.
Using \textbf{Approach 1} (Definition~\ref{def:evenlog1}) there may be two events having the same value, i.e., $e=(c,a,t,r)$ does \textbf{not uniquely identify} an event.
However, it is easy to create new event logs such as:
$L_1 = [(c,a,t+5,r) \mid (c,a,t,r) \in L]$ (all events were delayed by five time units), 
$L_2 = [(c,a,t,r) \mid (c,a,t,r) \in L \ \wedge \ a \not\in A ]$ (all $A$ events were removed), and
$L_3 = [(c,a,t,r) \mid (c,a,t,r) \in L \ \wedge \ r = \mi{John} ]$ (only the activities conducted by John are retained). 

Using \textbf{Approach 2} (Definition~\ref{def:evenlog2}), we can uniquely identify events. In any context, 
$e \in \Uev$ is a specific event having \textbf{immutable properties} such as 
$\pi_{case}(e)$, $\pi_{act}(e)$, $\pi_{time}(e)$, and $\pi_{res}(e)$.
This is an advantage and also a disadvantage. There is no need to introduce $\Uev$ and $\pi$, because they exist universally. 
However, it is impossible to change event attributes, e.g., add a delay.
It is incorrect to say $\pi_{time}(e) = \pi_{time}(e)+5$, because the attribute values are fixed.

\textbf{Approach 3} (Definition~\ref{def:evenlog3}) can be seen as a combination of the above approaches.
Given an event log $L = (E,\pi)$, an event $e \in E$ can be uniquely identified. However,
there may be two distinguishable events with the same attribute values, e.g., $e_1,e_2 \in E$ and $\pi(e_1) = \pi(e_2)$.
Moreover, it is possible to create new event logs using preexisting event logs.
Let $L = (E,\pi)$ and $L' = (E',\pi')$ such that 
$E' = \{e \in E \mid \pi(e)(\mi{act}) \not\in A \ \wedge \ \pi(e)(\mi{res}) = \mi{John}\}$ and
$\pi' \in  E' \rightarrow \Umap$, $\pi'(e) = \pi(e) \oplus \{(\mi{time},\pi(e)(\mi{time})+5),(\mi{res},\mi{Mary}),(\mi{costs},10)\}$ for $e \in E'$.
Note that for $L'$ the $A$ events are removed and only the events conducted by $\mi{John}$ are kept.
The timestamp of each remaining event is increased, the resource is changed, and a cost attribute is added.

\textbf{Which of the three approaches is most suitable, depends on your goal.}
However, do not use Approach 1 if you want to point to a specific event
and do not use Approach 2 if you want to create event logs from other event logs.
In all cases, you need to introduce event logs in definitions, lemmata, theorems, etc. 
Always start with ``Let $L = \ldots$'' to fix the context. 

\section{Avoid Pseudo-Code, Notations in Running Text, and Repetitions}
\label{sec:pseudo}
 
The author prefers to avoid using pseudo-code as much as possible.
In most cases, we want to explain an approach or present a novel idea, i.e., the focus is on ``What'' rather than ``How''.
We rarely want to discuss implementation details.
Of course, there are exceptions, e.g., to prove the complexity of an algorithm.
When it is possible to formalize things in a few lines, it is better to avoid pseudo-code. 
Pseudo-code is often ambiguous and non-declarative. 

\begin{figure}[thb!]
\begin{pseudocode}{MergeSort}{\sigma}
\label{MergeSort}
n \GETS \card{\sigma}\\
\IF n=2 \THEN
\BEGIN
\IF \sigma[1]>\sigma[2] \THEN
\BEGIN
x \GETS \sigma[1]\\
\sigma[1]\GETS \sigma[2]\\
\sigma[2]\GETS x
\END
\END
\ELSEIF n>2 \THEN
\BEGIN
m\GETS \lfloor n/2 \rfloor\\
\sigma' \GETS \la~\ra\\
\FOR i\GETS 1 \TO m \DO \sigma' \GETS \sigma' \cdot \la \sigma[i] \ra\\
\CALL{MergeSort}{\sigma'}\\
\sigma'' \GETS \la~\ra\\
\FOR j\GETS m+1 \TO n\DO \sigma'' \GETS \sigma'' \cdot \la  \sigma[j] \ra\\
\CALL{MergeSort}{\sigma''}\\
i\GETS 1\\
j\GETS 1\\
\FOR k \GETS 1 \TO n \DO

\BEGIN
\IF (i  \leq  m \AND  j \leq  n-m \AND \sigma'[i] \leq \sigma''[j]) \OR j > n-m \THEN
\BEGIN
\sigma[k]\GETS \sigma'[i]\\
i\GETS i+1
\END
\ELSE
\BEGIN
\sigma[k]\GETS \sigma''[j]\\
j\GETS j+1
\END
\END
\END\\
\RETURN \sigma
\end{pseudocode}
\caption{Example pseudocode.}
\label{fig-pseudo}
\end{figure}

As an example, consider the pseudo-code example in Figure~\ref{fig-pseudo}.
The pseudo-code is detailed and precise. However, if the goal is to sort a sequence, you can also state this in a compact, more declarative, manner.

\begin{definition}[Sorting]\label{def:sort}
Function $\mi{sort} \in \mathbb{R}^* \rightarrow \mathbb{R}^*$ is such that 
for any $\sigma = \la x_1, x_2,\allowbreak \ldots,\allowbreak  x_n\ra$:
$\mi{sort}(\sigma) = \la y_1, y_2, \ldots, y_n\ra$ with $[x_1, x_2, \ldots,\allowbreak x_n] \allowbreak = [y_1, y_2, \ldots, y_n]$
and $y_i \leq y_{i+1}$ for $1 \leq i < n$.
\end{definition}

This example looks far-fetched. However, this is what happens in many papers.
If the goal is not to formally reason about the complexity of an algorithm, but to present an approach, technique, idea, 
then describe things in a declarative manner.
Pseudo-code (often  disconnected from earlier definitions) can often be described 
more precisely in just a few lines of mathematics.
The Alpha algorithm can be fully defined in just eight lines of basic mathematics. Pseudo-code would make it imprecise and much longer.

You do not need nested loops when you can use $\forall_{x \in X} \ldots$, $\exists_{x \in X} \ldots$, and $\sum_{x \in X} \ldots$.
Also note that \textbf{bijections} are surprisingly powerful,
e.g., $\mi{sort}(\la x_1, x_2,\allowbreak \ldots,\allowbreak  x_n\ra) = \la y_1, y_2, \ldots, y_n\ra$ implies 
that there is a bijective function $f \in \{1,\ldots ,n\} \rightarrow \{1,\ldots ,n\}$ such that
$x_i = y_{f(i)}$ and $y_i = x_{f^{-1}(i)}$ for any $1 \leq i \leq n$. 
Often, we need such one-to-one relationships.

\textbf{Do not define things (e.g., notations and concepts) in running text.} 
There are three main reasons for this: 
(1) the reader cannot find the place where the notation or concept was introduced when it is used later,
(2) the context is unclear (i.e., the ``Let ...'' part is missing), and
(3) it is often used to hide known gaps and shortcuts.
Of course, it is OK to use running text to ``refresh'' standard concepts like 
$N=(P,T,F)$, $M \in \bag(P)$, $\pre x$, $\post x$, etc.
However, often known concepts end up in nice explicit definitions, but the new concepts are defined in the running text.
This is not acceptable. The paper needs to be balanced. 
You cannot assume on page 2 that the reader is an idiot, and on page 6 assume that (s)he can read your mind and has super-powers.
If you lack space, decide what to leave out.
This provides the space to address comments by reviewers.

\textbf{Avoid pages with just plain text.} It makes the paper look verbose and few will read it.
Try to use \emph{italics} to emphasize things. Also, use itemized lists and tables to structure your ideas. If you can put it in a table or figure, do not use plain text. They say ``A picture is worth a thousand words'', so you will save space \smiley.

There is a tendency to \textbf{repeat arguments and vague informal sentences when things are not perfectly clear for the author}.
This will make things worse and is like ``rubbing a stain''.
Be aware of this when you try to ``clarify by repetition''. 

\section{Figures, Headings, Etc.}
\label{sec:figs}

Figures are very important to convey a message and to structure your ideas. When writing a paper, \textbf{start with your formal definitions and figures}. Do \textbf{not} start with the abstract and introduction.
Make sure figures are readable and self-contained. 
Note that text in font size 9 or smaller, cannot be read by older professors \smiley{}.
The \textbf{caption is very important} and should be extensive enough. Ask yourself: What should the reader know and conclude from this. 
Figures often end up on a different page, so you cannot assume that the reader is looking a the figure while reading the text explaining it.
\begin{figure}[thb!]
\centering
\includegraphics[width=10.0cm]{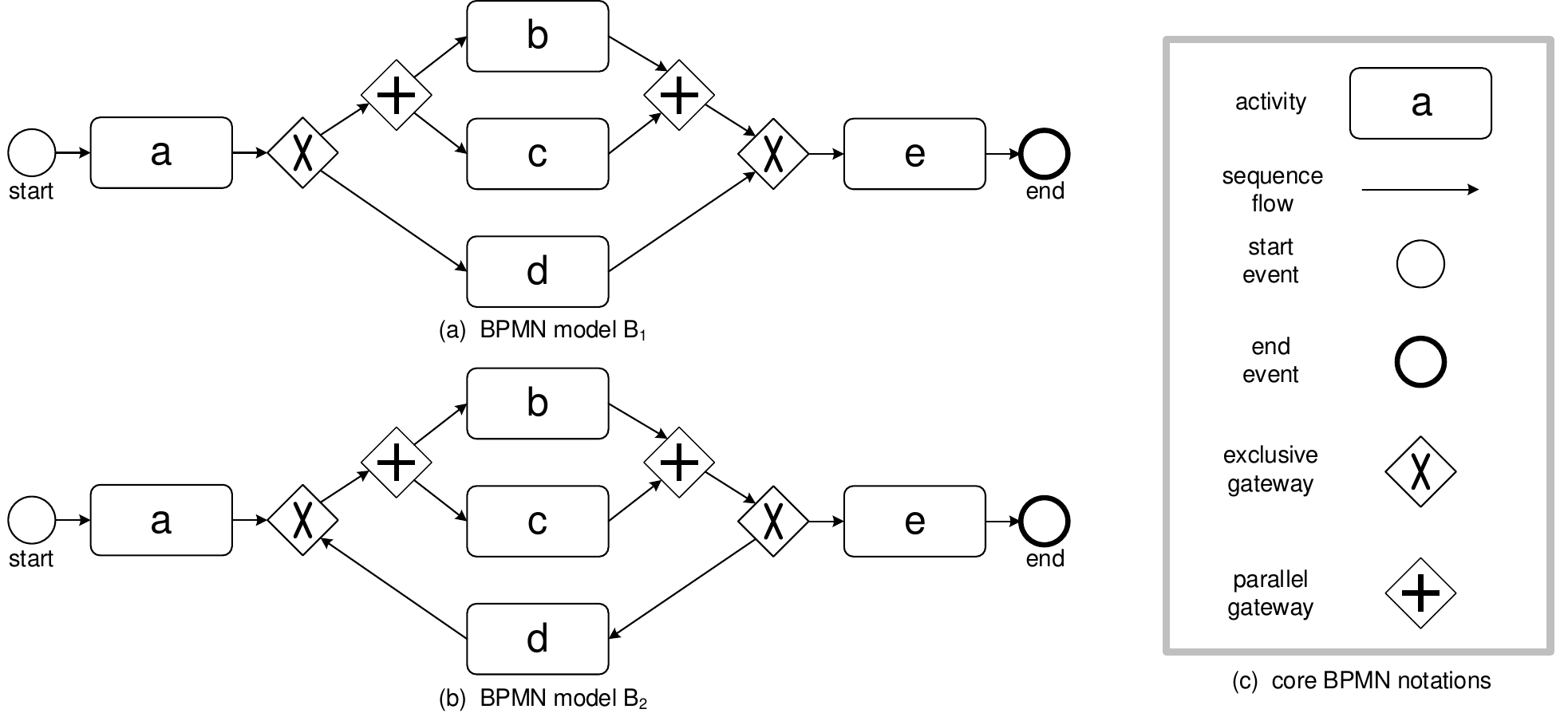}
\caption{Two BPMN models: $B_1$ and $B_2$ illustrating the core BPMN notations.
$\mi{lang}(B_1) = \{\la a,b,c,e\ra,\allowbreak \la a,c,b,e\ra,\allowbreak \la a,d,e\ra\}$ and 
$\mi{lang}(B_2) = \{\la a,b,c,e\ra,\allowbreak \la a,c,b,e\ra,\allowbreak \la a,b,c,d,b,c,e\ra,\allowbreak \la a,c,b,d,b,c,e\ra, \ldots \}$.
Note that $B_2$ has infinitely many accepting traces.}
\label{fig-BPMN-examples}
\end{figure}

Non-informative captions like ``Petri net $N$'' and ``Event log $L$'' do not help the reader.
It is also vital to have good captions and explain symbols in a box (like a map legend).
Use the space you have and add enough explanations.
See, for example, Figure~\ref{fig-BPMN-examples}. The caption and list of symbols help the reader to inspect and understand the BPMN notation.
Hardly any additional text is needed.

\textbf{Many papers that present experimental results use diagrams that can only be understood by the first author.}
Bad charts and graphs are omnipresent.
Use the right chart, e.g., do not use pie charts to compare numbers (use a barchart) 
and do not use bar charts to summarize continuous data.
Using standard RGB colors is a bad idea. Use tools like the \url{colorbrewer2.org} to ensure readability.
To see differences between values, the length of two lines or the direction of a line are easier to tell apart than shades of gray or the intensity of a color. Showing 3D shapes to show magnitude creates confusion: The height, surface, and volume of a 3D shape represent very different scales.

Visualization is a topic in itself. 
However, at the minimum, \textbf{explain the X- and Y-axes in a truly understandable manner}.
When plotting different lines, tell what they represent. Often acronyms are added that were explained in another section in running text.
\textbf{Use the caption and a visual list of symbols, algorithms, etc.}
The reader should understand what the diagram represents \textbf{without reading the text}. 
The text is there to explain phenomena and challenge hypotheses and not to explain things on a different page in a half-baked way.

\textbf{Carefully pick sections and subsections.}
A section can never contain just a single subsection.
The titles should be informative, and capitalization should be consistent. 
Check the style of the journal or conference and do this right from the start (not after someone else points out what you could see yourself).

\section{Typical LaTeX Problems}
\label{sec:latex}

Most papers in our field are written in LaTeX. This allows for consistent formatting and beautiful formal/mathematical expressions.
However, there are also recurring issues that inexperienced LaTeX users encounter.

\textbf{Overleaf encourages sloppiness} and leads to papers that can only be compiled in Overleaf.
Check out the error messages and make sure your LaTeX code is portable and still runs 5 years from now.
The LaTeX errors, BibTeX errors, and warnings for spelling errors are not there for decoration; repair all errors before sharing with others.
When working in a team, expect \textbf{Garbage-in Garbage-out} (GiGo): If you do not care, why should someone else?

Some LaTeX hints you may want to use (the colored text fragments are literal LaTeX statements):
\begin{itemize}
\item  After abbreviations correct spacing if needed. For example, write {\color{green}\verb|i.e., |},   {\color{green}\verb|e.g., |},   {\color{green}\verb|i.e.\ |},  or {\color{green}\verb|e.g.\ |}, but not
{\color{red}\verb|i.e. |} or {\color{red}\verb|e.g. |} After a normal period there is a larger space.
\item Write {\color{green}\verb|Figure~\ref{...}|}, {\color{green}\verb|Section~\ref{...}|}, etc. to avoid line breaking.
\item If needed use 
{\color{green}\verb|\usepackage{url}|},
{\color{green}\verb|\usepackage{amssymb}|},\\
{\color{green}\verb|\usepackage{amsmath}|},
{\color{green}\verb|\usepackage{enumitem}|},\\
{\color{green}\verb|\usepackage{graphicx}|}, etc., but disable if not needed.
\item Use shorthands, e.g., {\color{green}\verb|\def\la{\langle}|} and 
{\color{green}\verb|\def\ra{\rangle}|}. The symbols {\color{red}\verb|< ... > |} are wrong.
\item Add {\color{green}\verb|\def\mi#1{\mathit{#1}}|} at the top of the document. Use this to create identifiers in mathematics font that consist of more than one letter. For example, write \\
{\color{green}\verb|$\mi{filter} \in A \rightarrow B$|}
and not \\
 {\color{red}\verb|$filter \in A \rightarrow B$|}\\
See the differences between {\color{green}$\mi{filter} \in A \rightarrow B$} and {\color{red}$filter \in A \rightarrow B$}.
Some more examples: 
{\color{red}$ff(Node)$} versus {\color{green}$\mi{ff}(\mi{Node})$},
{\color{red}$XYZ$} versus {\color{green}$\mi{XYZ}$}, and\\
{\color{red}$Donaudampfschiffahrtsgesellschaftskapitaen$} versus\\ {\color{green}$\mi{Donaudampfschiffahrtsgesellschaftskapitaen}$}.
Yes, it is subtle, but everyone will notice it.
\item Ensure enough space in expressions. Compare\\ 
{\color{red}$\exists_{x \in X} b_1(x)=a_1(x) \wedge  b_2(x)=a_2(x)$} generated by\\ 
{\color{red}\verb|$\exists_{x \in X} b_1(x)=a_1(x) \wedge  b_2(x)=a_2(x)$|}
and\\
{\color{green}$\exists_{x \in X}\ b_1(x)=a_1(x) \ \wedge \ b_2(x)=a_2(x)$} generated by\\ 
{\color{green}\verb|$\exists_{x \in X}\ b_1(x)=a_1(x) \ \wedge \ b_2(x)=a_2(x)$|}.

\item Use {\color{green}\verb|\allowbreak|} to allow breaking lines in mathematical expressions. 
Add it to places in an expression where line breaks may make sense.
This avoids hard coding explicit line breaks that go wrong after you change the text or use a different style.
\end{itemize}

Note that these hints are a random sample. Also avoid using hard-coded references. Always use a symbolic reference, e.g. to refer to a figure ({\color{green}\verb|Figure~\ref{label}|}) or a section ({\color{green}\verb|Section~\ref{label}|}).
This way you can change the paper without renumbering.

\section{References}
\label{sec:refs}

Invest time to create a \textbf{good .bib file} with \textbf{complete} information using a \textbf{unified style}.
Some hints:
\begin{itemize}
\item Be consistent! For example, using capitals in titles or not, and using first names or just initials. \textbf{It looks very sloppy if you mix different styles.} Note that journals often require a particular style for the final version. However, already during the reviewing process, things should look uniform. You can adapt the style, but be consistent. Many .bib files are concatenations of different styles and authors, creating a bad impression.  
\item Do not use abbreviations in references unless you are forced to do so in the final version.
\item Add all information. Do not forget page numbers, editors, publisher, volume number, etc. 
Avoid adding extra information like a URL if you are not prepared to add this for all references.
\item You only lose time if you do not get it right the first time. Do not disrespect or insult the reader.
\end{itemize}

Try to \textbf{reduce the number of self-references}. If more than half of the references come from the same group, 
reviewers will find reasons to reject the paper. As a PhD there is no need to list all your papers (be very selective). 
Avoid the impression  that the paper is ``more of the same''.
Being in a larger successful group, this may not be so easy, especially for specialized topics.
However, you can always lift the abstraction level. Using Scopus and other tools, it is really easy to find related work that you were not aware of.
\textbf{Do a small systematic literature review and you will be surprised!}

If you submit a paper to a journal or conference, and you do not refer to any of the papers published in that journal or conference, then \textbf{expect a reject}. This may seem unfair. However, if you want to send a paper to a journal or conference in a neighboring field, 
you need to show that it fits. Note that journals try to improve their impact factors by encouraging authors to refer to papers in the journal.
This triggers many desk rejects.

Most of our papers also present software. 
Make sure that the reader/reviewer \textbf{believes the software exists and works as described}.
It is not enough to refer to a GitHub or your personal website.
Very few readers/reviewers will spend more than 10 minutes to get things running.
As stated before, \textbf{``You only get one chance to make a first impression''}.
Without an installer, UI, and professional website, \textbf{the first impression will be the last}.
Also imagine looking at the tool and website in 5-10 years from now.
If you do not care, why should the reader/reviewer care?

\section{Dare to Restart!}
\label{sec:dare}

Section~\ref{sec:start} provided a checklist with questions such as:
\begin{itemize}
\item What is the problem and why is the problem relevant?
\item What is your unique way of approaching the problem?
\item What is the input and what is the output?
\end{itemize}
These questions need to be answered \textbf{before} writing the paper. 
\textbf{It is strongly recommend to first present the story-line of the paper to colleagues in 10-15 minutes.} If this is impossible, do not start writing and first revisit the questions above.

Despite these efforts it can be happen that things that seemed clear at the beginning become less clear while writing. Often unanticipated complexities emerge when conducting experiments or detailing algorithms and proofs. \textbf{This often leads to ``space problems''.} A common mistake is that preliminaries take too much space and the later parts (e.g., experiments) are too brief. Also there is no point in formalizing the basics and be informal about the actual approach. Many papers formally define for example what a Petri net is, but not actually use the formalization. This relates to the earlier comment about making consistent assumptions about your audience  (Section~\ref{sec:start}).  Another common mistake is that the paper starts very broad (``process improvement'') and after five pages suddenly reduces the scope to a much smaller or more specific problem (``scheduling resources'').  \textbf{If you run into space problems, you probably need to restart from scratch.} In any case ensure that the paper is balanced and drop the things that cannot be explained properly or that are out-of-scope.

When a paper is reviewed for a workshop or conference, \textbf{reviewers typically ask for more} explanations, more experiments, more related work, etc. However, the page limit remains the same. Especially when a paper is rejected, do not try to squeeze more words into the paper in the hope to clarify things.
Adding more text without removing text will not make things clearer.
Remember Michelangelo's quote: \textbf{``I saw the Angel in the marble and carved until I set him free''} (i.e., ``less is more''). If the story-line is unclear or the notations cannot be understood by reviewers, then adding more ``stuff'' does not help.

\textbf{If you write the paper from scratch, you will notice that you are able to convey the same message in a clearer and more succinct manner.}
Of course this is time consuming, but sometimes you need to ``bite the bullet''.

\section{Conclusion}
\label{sec:conclusion}

This style guide aims to provide concrete suggestions and 
help authors to write better papers that can stand the test of time.
The goal of any researcher should be to write papers that have an impact and progress science.
This is only possible if papers are accessible and of good quality.

We used Dijkstra's proverb \textbf{Beauty Is Our Business} to set the ambition level for scientific papers.

As mentioned, some things are a matter of taste and some of the statements should be taken with a grain of salt.
The partly provocative statements are intended to make authors think about ``paper writing habits'' and their effect on readers and reviewers. 
Feedback and supplementary recommendations are welcome!

~\\
~\\{\bf Acknowledgments}\\ The author thanks the Alexander von Humboldt (AvH) Stiftung for supporting his research.

\bibliographystyle{plain}
%\bibliography{../../../bib/lit}
\bibliography{lit}

\end{document}